\documentclass[twocolumn]{aastex631}

\newcommand{\Msun}{{\rm M_{\odot}}}

\newcommand{\kpc}{\, {\rm kpc}}

\newcommand{\K}{\, {\rm K}}
\newcommand{\kmps}{\, {\rm km \, s^{-1}}}

\begin{document}

\title{On the Lifetime of Molecular Clouds with the ``Tuning-Fork" Analysis}
\shorttitle{Molecular Cloud Lifetime with the Tuning-Fork Analysis}
\shortauthors{Koda \& Tan}

\correspondingauthor{Jin Koda}
\email{jin.koda@stonybrook.edu}

\author[0000-0002-8762-7863]{Jin Koda}
\affil{Observatoire de Paris, LERMA, PSL Univ., CNRS, Sorbonne Univ., Paris, France}
\affil{Department of Physics and Astronomy, Stony Brook University, Stony Brook, NY 11794-3800, USA}

\author[0000-0002-3389-9142]{Jonathan C. Tan}
\affil{Dept. of Space, Earth \& Environment, Chalmers University of Technology, SE-412 96 Gothenburg, Sweden}
\affil{Department of Astronomy, University of Virginia, Charlottesville, VA 22904-4325, USA}

\begin{abstract}
The ``tuning-fork" (TF) analysis of CO and H$\alpha$ emission has been used to estimate the lifetimes of molecular clouds in nearby galaxies.
With simple model calculations, we show that this analysis does not necessarily estimate cloud lifetimes,
but instead captures a duration of the cloud evolutionary cycle, from dormant to star forming, and then back to a dormant phase.
We adopt a hypothetical setup in which molecular clouds (e.g., traced in CO) live forever and form stars (e.g., HII regions) at some frequency, which then drift away from the clouds.
The TF analysis still returns a timescale for the immortal clouds.
This model requires drifting motion to separate the newborn stars from the clouds, and we discuss its origin.
We also discuss the physical origin of the characteristic spatial separation term in the TF analysis
and a bias due to systematic error in the determination of the reference timescale. 
\end{abstract}


\section{Introduction}\label{sec:intro}

Recently, the ``tuning-fork" (TF) analysis 
has been applied to CO and H$\alpha$ emission maps
of nearby galaxies \citep{Schruba:2010fk, Kruijssen:2014aa, Kruijssen:2018aa}.
These analyses have been interpreted to indicate short lifetimes of molecular clouds
\citep[5-30~Myr; ][]{Kruijssen:2019aa, Chevance:2020ab, Chevance:2022aa, Kim:2022aa}.
A fundamental assumption for these conclusions is that molecular clouds die shortly after their first star formation (SF) events.
However, even if clouds live forever with occasional SF that then undergoes relative drift with respect to the clouds,
this method returns \textit{some} timescale, which is \textit{not} a cloud lifetime.
In this paper, we demonstrate this and reconsider the meaning and significance of this method.

The TF analysis is built on two earlier developments.
First, \citet{Kawamura:2009lr} analyzed molecular cloud populations in
the Large Magellanic Cloud (LMC).
They classified clouds according to SF phases
(clouds with no massive SF, with HII regions, and with HII regions and young clusters),
and counted the number and fraction in each phase.
Using a typical age of the star clusters as an absolute reference duration in the last phase,
they translated the population fractions into the durations
that clouds spend in each phase.
They assumed that clouds die after the Type III phase
and summed up the durations in Type I, II, and III to
estimate a cloud lifetime of 20-30~Myr.

Second, \citet{Schruba:2010fk} introduced a TF analysis to
measure a size scale beyond which local fluctuations
of CO and H$\alpha$ emission average out.
They plotted the CO/H$\alpha$ flux ratio with increasing aperture size.
They set CO or H$\alpha$ peaks at aperture centers,
which form the upper and lower branches of the TF, respectively.
The two branches show a large separation at small apertures,
but approach and merge toward larger apertures.
They took the aperture at the merging point as the ``average-out scale''
of $\sim 300$~pc in M33.

\citet[][hearafter, KL14]{Kruijssen:2014aa} and \citet[][hearafter, K18]{Kruijssen:2018aa}
combined these two approaches.
They showed that the two branches of the TF,
as they approach and merge from small to large scales,
carry the information of the physical size and timescales
involved in the SF process.
They derived equations to extract the size and timescales
from an observed TF diagram.
This method is intrinsically an extension of the population analysis \citep{Kawamura:2009lr},
although their equation used the CO/H$\alpha$ flux ratio,
that is, the flux-weighted number ratio of molecular clouds in different SF phases,
instead of the pure number ratio.
For an absolute scale in time, they used a typical lifetime of HII regions of $\sim 4.32$~Myr \citep[][roughly a main sequence lifetime of very massive $\sim50\Msun$ stars from \citealt{Schaller:1992aa}]{Haydon:2020ab},
instead of the age of stellar clusters.
Hence, their cloud/SF sequence is slightly different from the one
in \citet{Kawamura:2009lr}, and runs from clouds without HII regions,
with HII regions, and HII regions without clouds (Figure \ref{fig:timescales}A).
By applying their equation to CO and H$\alpha$ observations,
\citet{Kruijssen:2019aa}, \citet{Chevance:2020ab},
and \citet{Kim:2022aa} obtained timescales of 5-30~Myr,
which they interpreted as cloud lifetimes.

Such short molecular cloud lifetimes are not in alignment with some observations.
It is difficult to reform the clouds from the diffuse atomic ISM on short timescales to compensate the rapid cloud destruction \citep{Scoville:1979lg, Koda:2016aa}.
Second, molecular clouds with and without SF are unevenly distributed in galactic disks.
The clouds with SF are mostly along spiral arms, and those without SF are in the interarm regions \citep[e.g., ][]{Koda:2023aa}.
Given disk rotation timescales of an order of 200~Myr,
the clouds cannot move from the interarm regions to spiral arms within the suggested short lifetimes of 5-30~Myr.

In this paper, we demonstrate that the TF analysis
returns a finite ``cloud lifetime" even when molecular clouds live forever, as long as there is a relative drift velocity between the formed stars and their natal molecular clouds.
This analysis does not necessarily trace cloud lifetimes,
and so the interpretation of the results needs to be altered.

The TF analysis is applicable to any tracer of gas clouds and SF.
We use the term ``cloud", or occasionally ``gas", for gas cloud tracers.
In observations, the CO emission is often used to trace this component.
We also use ``star particle", or simply ``star", for SF tracers.
The H$\alpha$ emission (i.e., HII regions) is most commonly used for this component, 
but other tracers, such as young star clusters and UV emission, are also possible.

\section{Method} \label{sec:method}

We apply the TF analysis to two simplified cases (toy models):
(A) molecular clouds are mortal and have a \textit{finite} lifetime as assumed by K18,
and (B) they are immortal and live forever with an \textit{infinite} lifetime\footnote{In practice, the term ``immortal'' in this work means that the clouds are very long-lived compared to the other timescales in the evolution that we discuss here.}.
The clouds are distributed over a disk, occasionally form stars, and eject them into the inter-cloud space.
The stars drift away from the clouds, as is consistent with observations \citep[e.g.,][]{Leisawitz:1989aa, Peltonen:2023aa}, and die after their lifetime.
The drift will be discussed further in Sections \ref{sec:drift} (a case of no drift) and \ref{sec:discussion} (origin of the drift).

\begin{figure*}[h]
\plotone{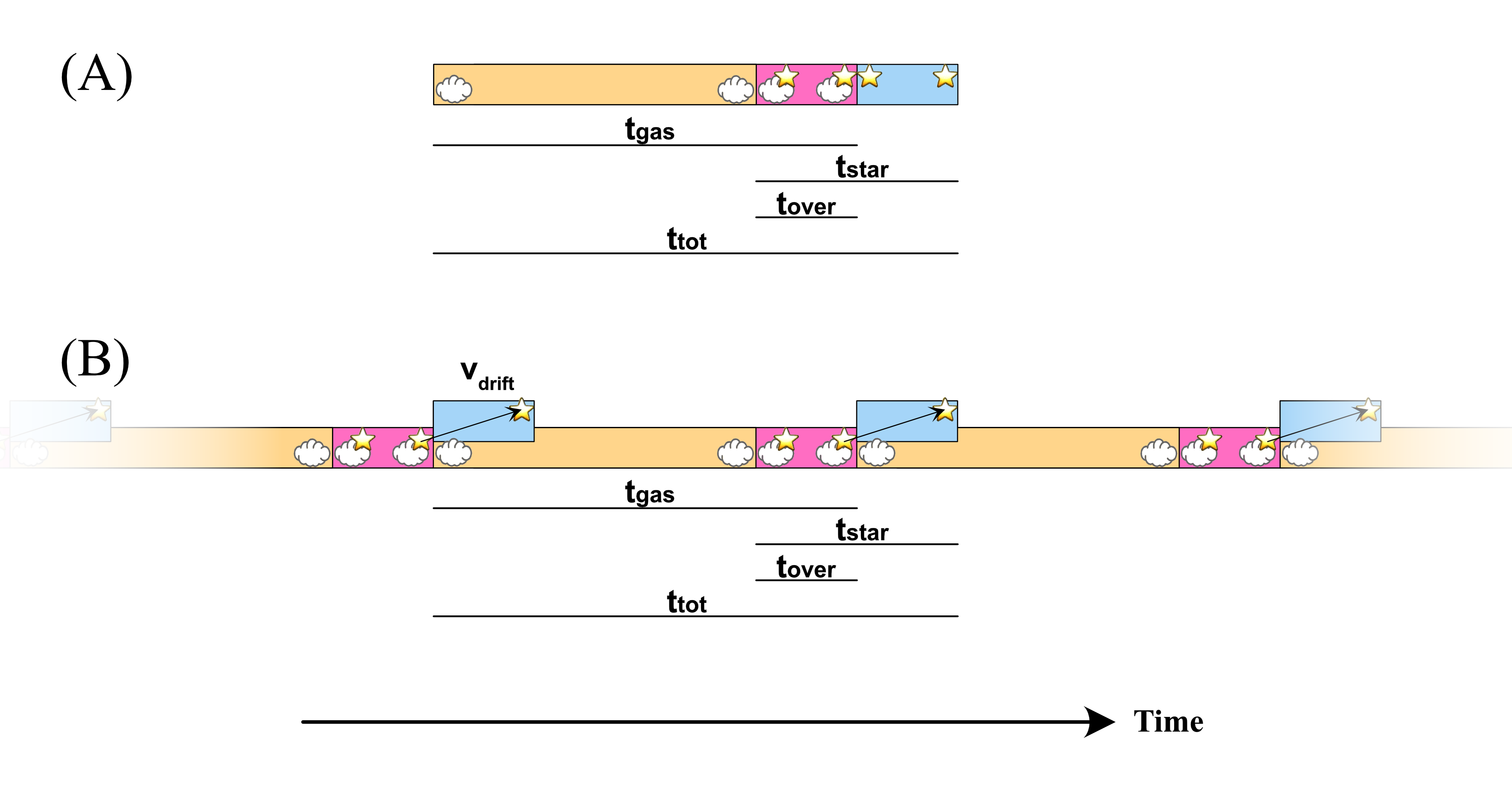}
\caption{Schematic presentations of
(A) mortal molecular clouds by \citet[][KL14]{Kruijssen:2014aa} and \citet[][K18]{Kruijssen:2018aa}, and
(B) immortal clouds with infinite lifetime.
In case A,
$t_{\rm gas}$ is the lifetime of gas cloud,
$t_{\rm over}$ is the timescale in which a cloud contains a star particle (SF tracer peak),
$t_{\rm star}$ is the lifetime of a star particle, and
$t_{\rm tot}$ is the total duration between the birth of a cloud and the death of its star particle.
In case B, clouds live forever and repeat a SF cycle.
The definitions of the timescales are the same as in case A, except for $t_{\rm gas}$, which is the timescale of one SF cycle in a gas cloud.
In both cases, $t_{\rm tot} = t_{\rm gas}+t_{\rm star}-t_{\rm over}$.
Note, case B involves a relative drift velocity, $v_{\rm drift}$, between stars and clouds.
\label{fig:timescales}}
\end{figure*}

\begin{figure*}[t]
\epsscale{1.0}
\plotone{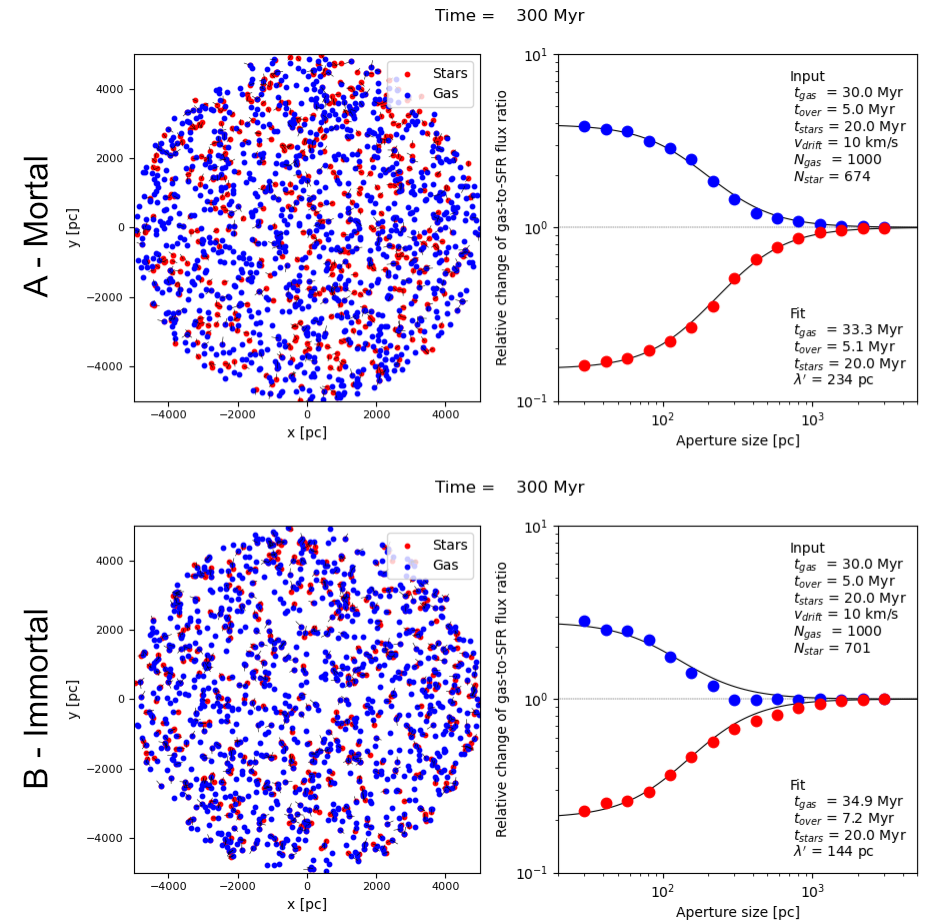}
\caption{
Snapshot of a randomly sampled uniform distribution of clouds at $t=300$~Myr.
\textit{Top:} Case A - mortal clouds and
\textit{Bottom:} Case B - immortal clouds.
The input parameters are shown in the right panels (top-right corners).
\textit{Left:} the gas/cloud distributions in blue.
They occasionally form star particles (red) with a lifetime of $t_{\rm star}$ at a probability of $p_{\rm SF}=\Delta t/t_{\rm gas}$ in each $\Delta t=$1~Myr timestep.
The gas clouds do not move from their birth locations.
The star particles stay/overlap with the clouds for $t_{\rm over}$, and then, drift away from the clouds and travel for the rest of their life.
Their drift velocity is $v_{\rm drift}$ in a random direction.
The black lines connect the start points (clouds) and end points of the stellar trajectories, but are short and often hidden behind the other markers.
In Case A (\textit{top}), clouds die/disappear at the end of the overlap period $t_{\rm over}$, and the same number of clouds
are redistributed uniformly over the disk (the total number $N_{\rm gas}$ stays constant).
In Case B  (\textit{bottom}), clouds live forever and stay at the same locations throughout the simulation.
\textit{Right:} the TF fits. The blue and red dots are from eq. (\ref{eq:bobs}) for gas clouds and stars, respectively.
The solid lines are fits of eqs. (\ref{eq:Bmodgas}) and (\ref{eq:Bmodstar}).
The fit results are also presented (bottom-right corners).
}
\label{fig:uni}
\end{figure*}

Figure \ref{fig:timescales} illustrates the lifecycle of molecular clouds
(A) for the mortal clouds, and (B) for the immortal clouds.
We simulate the evolution of the spatial distributions of the clouds and stars under each of these scenarios.
Figures \ref{fig:uni} and \ref{fig:expspi} (left) show example snapshots of the locations of clouds (blue) and star particles (red).
We apply the TF analysis to each snapshot.
The details of each setup are discussed in Section \ref{sec:results}.

Case A is designed to test the validity of our setup, i.e., in comparison to the previous results of KL14 and K18.
On the other hand, Case B shows that the TF analysis can be fit to the immortal clouds
and return a finite timescale, but which should not be interpreted as a cloud lifetime.

\subsection{The Observational Tuning-Fork Diagram} \label{sec:TFobs}

From the cloud and star distributions in each timestep,
we calculate the ``normalized relative gas-to-star flux ratio'', $B^{\rm obs}$, as a function of aperture size, $l$.
First, we draw a subsample of gas clouds (or star particles), $S_{\rm MC}$, using Monte Carlo realizations (see below)
and calculate $B^{\rm obs}$ for this subsample,
\begin{equation}
    B^{\rm obs, S_{\rm MC}}_{p} (l) =  \left( \frac{\Sigma_{i \in S_{\rm MC}} F^{\rm i}_{\rm gas}(l)}{\Sigma_{i \in S_{\rm MC}} F^{\rm i}_{\rm star}(l)} \right) / \left( \frac{F^{\rm tot}_{\rm gas}}{F^{\rm tot}_{\rm star}} \right) , \label{eq:bobsMC}
\end{equation}
where
$p$=gas or star (depending on which object type apertures are centered on).
Here $F^{\rm i}_{\rm gas}(l)$ and $F^{\rm i}_{\rm star}(l)$ are the gas and star fluxes within an aperture of diameter $l$ around an emission peak $i$, while
$F^{\rm tot}_{\rm gas}$ and $F^{\rm tot}_{\rm star}$ are the total gas and star fluxes in the disk
and are used to normalize the ratio.
The summations are taken over clouds (gas peaks) for the top branch of the TF plot (e.g., blue dots in Figures \ref{fig:uni} and \ref{fig:expspi} right panels) and over star peaks for the bottom branch (red dots).
We generate an $N_{\rm MC}$ set of subsamples \{$S_{\rm MC}$\} and average them to obtain
\begin{equation}
    B^{\rm obs}_{p} (l) = \frac{1}{N_{\rm MC}}\sum_{s \in \{S_{\rm MC}\}} B^{\rm obs, s}_{p}. \label{eq:bobs}
\end{equation}

K18 generated the subsamples to avoid aperture overlaps and double-counts of gas (or star).
Each subsample includes only the gas clouds (or star particles) whose apertures do not overlap each other.

Note that if all the gas peaks and star peaks have the same fluxes, respectively, the ratio of the summations in eq. (\ref{eq:bobsMC}) give, e.g., the population of clouds that have associated SF (the top branch).
In this sense, the TF analysis is similar to the population analysis by \citet{Kawamura:2009lr}.

\subsection{The Model Tuning-Fork Equations} \label{sec:TFmodel}

Figure \ref{fig:timescales}A shows the lifecycle of mortal clouds (Case A).
KL14 and K18 assumed that clouds are born and live for a duration of $t_{\rm gas}$.
Within their finite lifetime,
they form star particles (star clusters, HII regions, etc),
live with them for a duration of $t_{\rm over}$, and are dispersed (presumably) by stellar feedback.
The lifetime of star particles is $t_{\rm star}$ including $t_{\rm over}$.
The total cycle from cloud birth to stellar death is $t_{\rm tot}$,
and $t_{\rm tot}$ $=t_{\rm gas}+t_{\rm star}-t_{\rm over}$.
The cloud and star particles have constant fluxes, $f_{\rm gas}$ and $f_{\rm star}$, respectively.
With this picture, KL14 and K18 derived their TF fitting equations, by evaluating the relative gas-to-star flux ratio in an aperture of size $l$,
\begin{equation}
B^{\rm mod}_{\rm gas}(l)
    = \frac{F_{\rm star}^{\rm tot}}{F_{\rm gas}^{\rm tot}} \frac{F_{\rm gas}^{\rm ap}(l)}{F_{\rm star}^{\rm ap}(l)}.
\end{equation}

They split a galactic disk into many small areas of size $\lambda$
and assumed that they are independent of each other in terms of SF.
Within an aperture of size $l$, the number of the small areas is $N_{\lambda} \approx (l/\lambda)^2$.
If the fraction of the $N_{\lambda}$ areas that contribute to SF activity (either by having gas or stars) is $\eta$,
the number of the areas that would have a cloud is $\eta N_{\lambda} (t_{\rm gas}/t_{\rm tot})$.
When the aperture is intentionally centered on a cloud,
the number is $1+\eta N_{\lambda} (t_{\rm gas}/t_{\rm tot})$.
A statistical average of aperture flux centered on gas clouds is then
\begin{equation}
    F_{\rm gas}^{\rm ap}(l)= f_{\rm gas} \left[ 1+\eta N_{\lambda}\left( \frac{t_{\rm gas}}{t_{\rm tot}} \right) \right].
\end{equation}
The central cloud would have a star at a probability of $t_{\rm over}/t_{\rm gas}$, and its statistical average flux is $f_{\rm star} (t_{\rm over}/t_{\rm gas})$.
The number of the other stars in this aperture is $\eta N_{\lambda} (t_{\rm star}/t_{\rm tot})$.
Hence, the statistical average of stellar aperture flux in this case is
\begin{equation}
    F_{\rm star}^{\rm ap}(l)= f_{\rm star} \left[ \left( \frac{t_{\rm over}}{t_{\rm gas}} \right) + \eta N_{\lambda} \left( \frac{t_{\rm star}}{t_{\rm tot}} \right) \right].
\end{equation}

If the galactic disk has $N_{\rm tot}$ small areas of size $\lambda$, then
the numbers of clouds and star particles over the disk are $\eta N_{\rm tot} (t_{\rm gas}/t_{\rm tot})$ and $\eta N_{\rm tot} (t_{\rm star}/t_{\rm tot})$,
and their fluxes are $F_{\rm gas}^{\rm tot} = f_{\rm gas} \eta N_{\rm tot} (t_{\rm gas}/t_{\rm tot})$ and $F_{\rm star}^{\rm tot} = f_{\rm star} \eta N_{\rm tot} (t_{\rm star}/t_{\rm tot})$, respectively.
Hence, the relative gas-to-star flux ratio in an aperture of size $l$ centered at gas cloud is
\begin{equation}
    B^{\rm mod}_{\rm gas}(l)
    = \frac{1 + (t_{\rm gas}/t_{\rm tot}) ( l/ \lambda^{\prime} )^2}{ (t_{\rm over}/t_{\rm star}) + (t_{\rm gas}/ t_{\rm tot}) ( l/ \lambda^{\prime} )^2},  \label{eq:Bmodgas}
\end{equation}
where $\lambda^{\prime}\equiv\eta^{-1/2}\lambda$.

From a similar consideration,
the relative gas-to-star flux ratio in an aperture of size $l$ centered at star particle is
\begin{equation}
    B^{\rm mod}_{\rm star}(l)
        = \frac{ (t_{\rm over}/t_{\rm gas})+ (t_{\rm star}/t_{\rm tot}) ( l/ \lambda^{\prime} )^2 }{1 + ( t_{\rm star} / t_{\rm tot} ) ( l/ \lambda^{\prime} )^2 }.\label{eq:Bmodstar}
\end{equation}

These equations are the same as eqs. (15) and (16)
of KL14,
but are a simplified version by setting their flux evolution terms $\beta_{\rm s}=1$ and $\beta_{\rm g}=1$
since the fluxes are assumed to be constant in our work.
K18 also took into account
spatial extents of the clouds and star particles (in their eqs. 81 and 82),
but our particles are, for simplicity, assumed to be points without size.
We also note that KL14 and K18 did not account for the $\eta$ term, and their $\lambda$ is equal to $\lambda^{\prime}$ in this work.
We will discuss these parameters and their implications on estimating the physical scale of SF in Section \ref{sec:lambda}.

\subsubsection{Fit}

The observed TF diagram is fit with eqs. (\ref{eq:Bmodgas}) and (\ref{eq:Bmodstar}).
There are only three free parameters: $t_{\rm gas}$, $t_{\rm over}$, and $\lambda^{\prime}$.
We assume $t_{\rm star}$ is known.
We run a grid search in the three dimensional space to minimize
\begin{eqnarray}
    \chi^2&=&\sum_{l} \left[ (\log B^{\rm obs}_{\rm gas}(l)-\log B^{\rm mod}_{\rm gas}(l))^2 \right. \nonumber \\
        & & \left. + (\log B^{\rm obs}_{\rm star}(l)- \log B^{\rm mod}_{\rm star}(l))^2 \right].    
\end{eqnarray}

\section{Results} \label{sec:results}

We apply the TF analysis to mortal clouds (Case A; Figure \ref{fig:timescales}A) and immortal clouds (Case B; Figure \ref{fig:timescales}B)
for a uniform cloud distribution (Section \ref{sec:uniform}) and exponential disk plus spiral arm distribution (Section \ref{sec:expspi}).
The Case A examples in the following sections are given to confirm the validity of our toy setups for the tests of the TF analysis.
The Case B examples demonstrate that the TF analysis returns a finite ``cloud lifetime" for the immortal clouds.

\subsection{Uniform Distribution}  \label{sec:uniform}

Figure \ref{fig:uni} (left) show snapshots of this setup for mortal (top) and immortal clouds (bottom).
The molecular clouds (blue points) are uniformly distributed over a galactic disk with a radius of $R_{\rm gal}=5$~kpc.
The number of clouds is $N_{\rm gas}=1000$.
The simulations are run from $t=0$ to 500~Myr with a timestep of $\Delta t=1$~Myr.
The clouds are fixed at the birth locations and do not move spatially.

We adopt $t_{\rm gas}=30$~Myr.
This parameter has different physical meanings in Case A and B.
It is an average cloud lifetime in Case A (mortal clouds), and a timescale of the SF cycle in Case B (immortal clouds).

The clouds form star particles at a probability of $p_{\rm SF}=\Delta t/t_{\rm gas}=1/30$ in each timestep.
The star particles stay on top of the parental clouds for an overlap period of $t_{\rm over}=5$~Myr \citep{Peltonen:2023aa}.
Then, they drift away at a speed of $v_{\rm drift}=10\kmps$ \citep{Leisawitz:1989aa, Peltonen:2023aa}.
The star particles die after their lifetime of $t_{\rm star}=20$~Myr (including $t_{\rm over}$),
which is about a main sequence lifetime of $\sim 12\:\Msun$ stars
\citep[e.g.,][]{Schaller:1992aa}
[Note that the star particles represent multiple stars, similar to star clusters or associations. The observed age spread in OB associations is a few to a few tens of Myr \citep{Wright:2023aa}].
The cloud and star particles are assumed to maintain constant fluxes, $f_{\rm gas}$ and $f_{\rm star}$, respectively.

The clouds die in Case A, but otherwise do not lose any mass by SF for the duration of simulations (500~Myr).
This assumption is justified for the long gas consumption time of $\sim$2~Gyr in observed galaxies \citep[e.g.,][]{Bigiel:2011uq}.

\subsubsection{Mortal Clouds (Case A)} \label{sec:caseAuni}

Figure \ref{fig:uni} (top) shows a snapshot at $t=300$~Myr.
The initial set of clouds have a uniform age distribution between 0 and 30~Myr ($=t_{\rm gas}$).
The clouds form star particles and die/disappear after the overlap period.
The same number of new clouds are replenished and distributed uniformly over the disk.
Thus, $N_{\rm gas}=1000$ remains constant.
The drift directions of the star particles are set randomly.

After about one cycle $\sim t_{\rm tot}$, the system reaches an equilibrium.
Statistically,
we expect $N_{\rm star}=(t_{\rm star}/t_{\rm gas}) N_{\rm gas}\sim 667$.
Figure \ref{fig:uni} (top) shows $N_{\rm star}=674$, which is consistent with the expectation.

We calculate the TF diagram for each snapshot (eq. \ref{eq:bobs}) and fit the TF equations (eqs. \ref{eq:Bmodgas} and \ref{eq:Bmodstar}).
We adopt $N_{\rm MC}=100$ (Section \ref{sec:TFobs}).
The top-right panel of Figure~\ref{fig:uni} shows a TF pattern and fitted lines for the snapshot.
This panel also shows the input parameters (top-right corner) and fit results (bottom-right).
The results of ($t_{\rm gas}$, $t_{\rm over}$)= (33.3, 5.1)~Myr are close to the input values of (30.0, 5.0)~Myr.
Therefore, our simplified setup is viable to test the TF analysis.

The derived $\lambda^{\prime}=234$~pc is close to the maximum star drift distance of 230~pc
[$=v_{\rm drift} (t_{\rm star}-t_{\rm over})$],
but also to a typical separation of 228~pc between regions involved in SF activity (having either gas or star).
In Section \ref{sec:discussion}, the meaning of $\lambda^{\prime}$ will be discussed:
unlike ($t_{\rm gas}$, $t_{\rm over}$), 
$\lambda^{\prime}$ depends on the global gas distribution, not only on the local properties of SF activity (such as drift distance).
To estimate the separation mentioned above, we start from $N_{\rm gas}=1000$ and $N_{\rm star}=$674 in this snapshot.
The number of gas cloud and star particles that overlap on top of each other is $\sim$168,
by averaging two estimations: $N_{\rm gas} (t_{\rm over}/t_{\rm gas}) \sim $167 and $N_{\rm star} (t_{\rm over}/t_{\rm star}) \sim $169.
Hence, the number of independent regions involved in SF is $N_{\rm ind}$=1506 (=1000+674-168).
An expected typical separation is $(\pi R_{\rm gas}^2/N_{\rm ind})^{1/2}\sim$ 228~pc.

\subsubsection{Immortal Clouds (Case B)} \label{sec:caseBuni}

The setup is the same as that in Section \ref{sec:caseAuni}, except that the clouds live forever (immortal).
Thus, the clouds do not disappear nor are replenished.
For simplicity we assume the clouds stay at their initial locations throughout the simulation and keep forming star particles at the same probability $p_{\rm SF}$.

The bottom-left panel of Figure \ref{fig:uni} shows a snapshot at $t=300$~Myr.
The cloud and star distributions result in a reasonable TF diagram (the bottom-right panel).
The TF fit provides ($t_{\rm gas}$, $t_{\rm over}$)= (34.9, 7.2)~Myr, which is close to the input values of (30.0, 5.0)~Myr.
There are slight deviations between the data and fits in the TF diagram.
Such slight deviations have been seen in previous observational analyses (e.g., Figure 2 in \citealt{Chevance:2020aa} and Figure 1 in \citealt{Kim:2022aa}),
and are not important for the main discussion in this work.

Therefore, we conclude that the TF analysis returns a result even when its basic assumption for cloud mortality is invalid.
The $t_{\rm gas}$ returned by the fit is not a cloud lifetime, since the clouds are immortal by definition, but is rather related to a timescale of the SF cycle within the immortal clouds.

\subsection{Exponential Disk + Spiral Arm Distribution} \label{sec:expspi}

We use the same setup as in Section \ref{sec:uniform}, but adopt
the gas distribution following a radial exponential disk profile plus two-arm spiral pattern.
For the distribution, we draw random numbers by combining the two probability distribution functions (PDFs).
For the exponential disk with a scale length of $h$, the PDF is
\begin{equation}
    P_{\rm exp}(r) = \exp \left(-\frac{r}{h} \right),
\end{equation}
in a cylindrical coordinate.
For a two-arm spiral pattern,
\begin{equation}
    P_{\rm spiral}(r, \theta) = P_{\rm exp}(r) \left( \cos \left[ \theta + p \ln \left(\frac{r}{h} \right) \right] \right)^{\gamma},
\end{equation}
where $p=\cot(i)$ with the spiral pitch angle $i$, and
\begin{equation}
    \gamma = \left\{
    \begin{array}{cc}
         1, & \text{for $r<w/\cos^{-1} (1/2)$} \\
         - \ln 2/ \left[\ln \left( \cos (w/r) \right) \right], & \text{otherwise.}
    \end{array}
     \right.
\end{equation}
This $P_{\rm spiral}$ roughly maintains the width of the spiral arm beyond $r=w/\cos^{-1} (1/2)$.
We adopt $R_{\rm gas}=5\kpc$, $h=3\kpc$, $w=0.6\kpc$, and $i=30\deg$.

We distribute 400 clouds in the exponential disk and 600 in the spiral arms with the total of $N_{\rm gas}=1000$.
Figure \ref{fig:expspi} (left) shows snapshots at 300~Myr.
The drift directions of the star particles are not random, but follow the clock-wise circular rotation.
The other parameters are the same as those in Section \ref{sec:uniform}.

\subsubsection{Mortal Clouds (Case A)} \label{sec:caseAexpspi}

Figure \ref{fig:expspi} (top) shows the case of mortal clouds at a snapshot of $t=300$~Myr.
The clouds die after the overlap period, and the same number of clouds are replenished with the same PDFs.

The TF analysis results in ($t_{\rm gas}$, $t_{\rm over}$)= (33.0, 5.5)~Myr, which is again close to the input values of (30.0, 5.0)~Myr.
This example shows that the ($t_{\rm gas}$, $t_{\rm over}$) estimates from the TF analysis are insensitive to the global gas distribution,
but capture the timescales relevant for the local physical processes for SF.

The derived $\lambda^{\prime}=183$~pc is smaller than the value derived for the uniform distribution (=234~pc in Section \ref{sec:caseAuni}),
though $N_{\rm gas}$ and $N_{\rm star}$ are very similar.
Therefore, $\lambda^{\prime}$ depends on the global distribution and is not a measure of the local physical processes for SF.

\subsubsection{Immortal Clouds (Case B)} \label{sec:caseBexpspi}

\begin{figure*}[t]
\epsscale{1.0}
\plotone{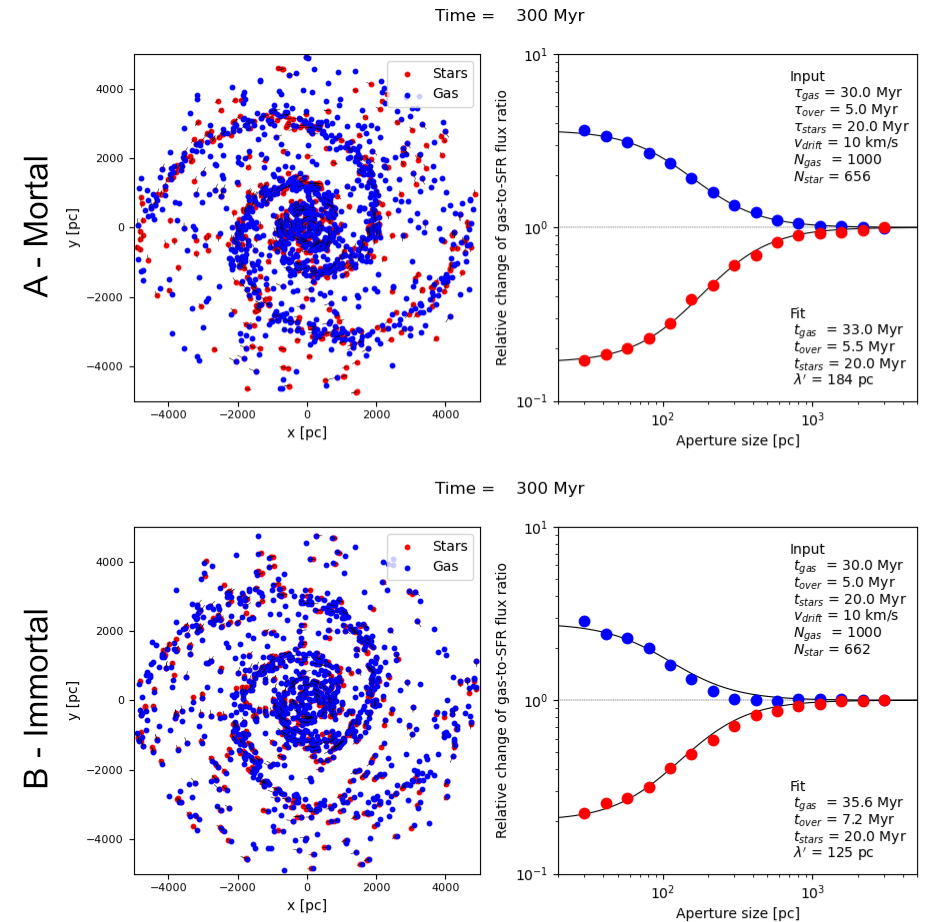}
\caption{
The same as Figure \ref{fig:uni}, but for the exponential disk plus spiral arm distribution.
We set the disk radius of $R_{\rm gas}=5\kpc$, scale length of the exponential disk of $h=3\kpc$, spiral arm width parameter of $w=0.6\kpc$, and pitch angle of $i=30\deg$.
As in Figure \ref{fig:uni}, the immortal clouds with occasional SF produce a TF pattern if the stars drift away from the parental clouds (bottom-right panel).
In this case the TF fit returns $t_{\rm gas}$, which is \textit{not} a cloud lifetime.
}
\label{fig:expspi}
\end{figure*}

Figure \ref{fig:expspi} (bottom) shows that 
immortal clouds give very similar results as for mortal clouds (Section \ref{sec:caseAexpspi}).
Even when the clouds live forever, the cloud and star distributions produce a TF pattern,
and the TF fit results in ($t_{\rm gas}$, $t_{\rm over}$)= (35.6, 7.2)~Myr (see Figure \ref{fig:expspi} bottom-right).
Again, the TF analysis returns a timescale even when the clouds are immortal, with this 
$t_{\rm gas}$ being related to the timescale of the SF cycle in the immortal clouds, rather than a cloud lifetime.

\begin{figure}[t]
\epsscale{1.0}
\plotone{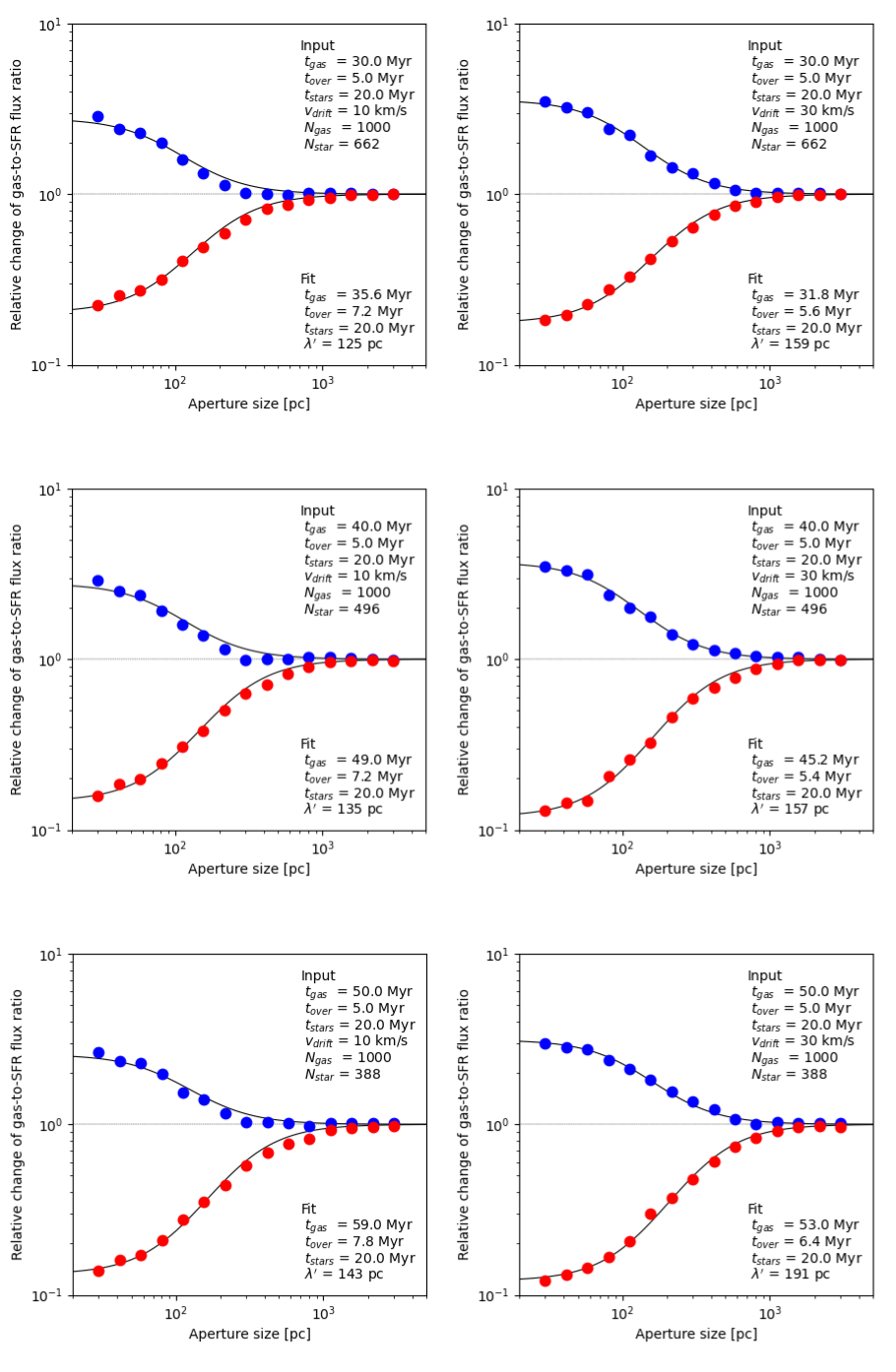}
\caption{
TF plots of immortal clouds.
The same as the bottom-right panel (immortal clouds) of Figure \ref{fig:expspi},
but for different values of $v_{\rm drift}$ and $t_{\rm gas}$:
$v_{\rm drift}=10\kmps$ (\textit{left}) and $30\kmps$ (\textit{right}).
$t_{\rm gas}=30$~Myr (\textit{top}),
$40$~Myr (\textit{middle}), and 
$50$~Myr (\textit{bottom}).
}
\label{fig:comb_expspi}
\end{figure}

Figure \ref{fig:comb_expspi} is the same as Figure \ref{fig:expspi} (bottom-right), but for
$v_{\rm drift}=$10 and 30$\kmps$ (columns) and $t_{\rm gas}=$30, 40, and 50~Myr (rows).
In all cases,
the fitted $t_{\rm gas}$ and $t_{\rm over}$ are close to the input values.

\subsection{Stellar Drift} \label{sec:drift}

\begin{figure}[t]
\epsscale{1.0}
\plotone{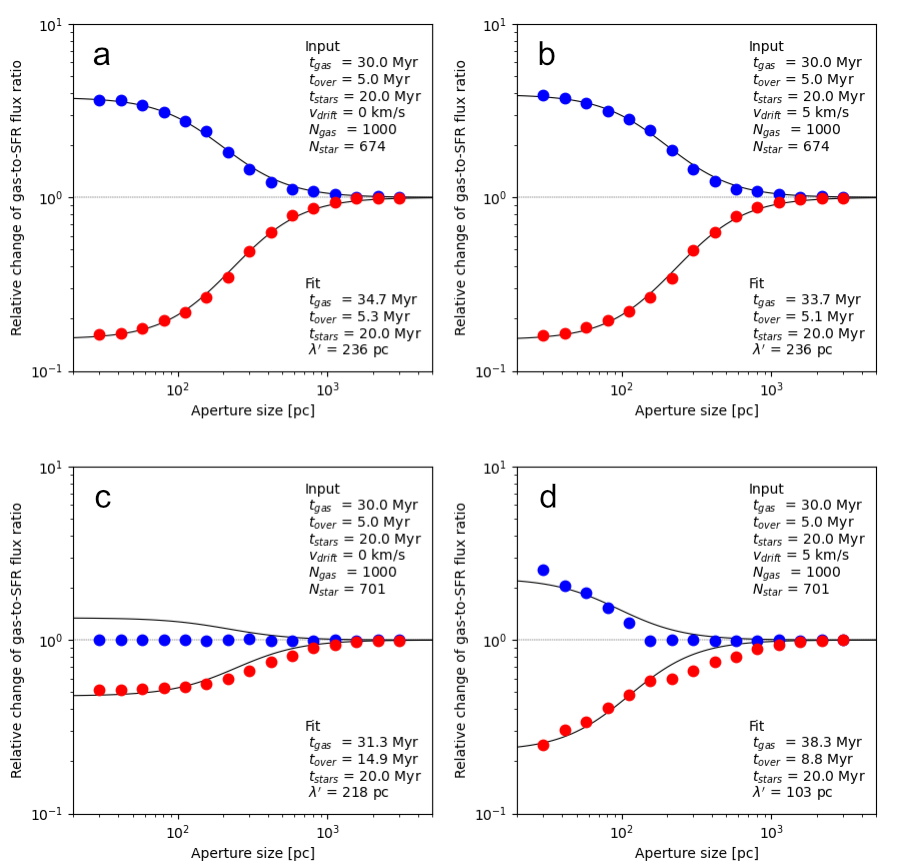}
\caption{
The same as Figure \ref{fig:uni} (uniform distribution), except for (\textit{Left}) $v_{\rm drift}=0\kmps$ and (\textit{Right}) $v_{\rm drift}=5\kmps$.
(\textit{Top}) mortal clouds and (\textit{Bottom}) immortal clouds.
In the case of $v_{\rm drift}=0\kmps$ (no drift), the stars do not drift away from the parental clouds. Hence, there are no stars without overlapping parental clouds when the clouds are immortal.
Thus, the top branch (blue) of the TF pattern stays at a value close to unity in panel (c).
}
\label{fig:nodrift}
\end{figure}

Stellar drift with respect to natal clouds is a key assumption in this work.
To see the importance of $v_{\rm drift}$,
we adopt the same parameters as in Section \ref{sec:uniform}, but set  $v_{\rm drift}=0\kmps$ and $5\kmps$.
Figure \ref{fig:nodrift} (a-d) show the TF diagrams for (left) $v_{\rm drift}=0\kmps$ and (right) $5\kmps$, and for (top: Case A) mortal clouds and (bottom: Case B) immortal clouds.

The mortal clouds (Case A) die after SF (after $t_{\rm over}$), and hence, there are star particles that do not spatially overlap with clouds.
This situation can produce a TF pattern even when they do not move away from the locations of the parental clouds.
In this model, the maximum star drift distance ($=v_{\rm drift} (t_{\rm star}-t_{\rm over})$) is the only size scale that characterizes the local properties of SF activity.
It varies as 0, 76, and 153~pc for $v_{\rm drift}=$0, 5, and 10$\kmps$, respectively.
However, their $\lambda^{\prime}$ is almost the same (234, 236, and 236~pc, respectively; see the top panels of Figures \ref{fig:uni} \& \ref{fig:nodrift}).
Thus, the $\lambda^{\prime}$ from the TF fit does not necesarily represent the local SF scale.

The immortal clouds (Case B) do not produce a TF pattern when $v_{\rm drift}=0\kmps$.
In Figure \ref{fig:nodrift}c, 
the top branch (blue points) stays constant at a value close to unity.
Since stars do not move away from the immortal clouds, all stars overlap with their parental clouds,
which makes the top branch flat.
This example shows that the clouds and stars have to be spatially separated to form a TF pattern.
Thus, if the clouds are immortal, the stars must drift away from the clouds to produce a TF pattern.

Figure \ref{fig:nodrift}d shows the case for $v_{\rm drift}=5\kmps$ for immortal clouds (Case B).
The maximum star drift distance is 76~pc.
Hence, when an aperture size is larger than $\sim$76~pc, all stars coexist with their parental clouds within the aperture.
This situation is the same as the one explained above for $v_{\rm drift}=0\kmps$,
and thus, the top branch is flat at large aperture size.
The apertures smaller than this distance can form a TF pattern as in this figure.

\subsection{Star Formation Probability with Galactic Rotation} \label{sec:sfprob}

We adopt the same parameters as in Section \ref{sec:caseBuni} (uniform distribution, Case B immortal clouds),
except for the SF probability.
We assume that the SF timescale is regulated by galactic rotation (dynamics).
We use $p_{\rm SF} = \Delta t / (b_{\rm rot} t_{\rm rot})$ [i.e., $t_{\rm gas}=b_{\rm rot} t_{\rm rot}$], where $t_{\rm rot} = 2 \pi R/V_{\rm rot}$ is the timescale of galactic rotation and $b_{\rm rot}$ is related to the efficiency of the SF process.
We adopt $V_{\rm rot}=200\kmps$ and $b_{\rm rot}=$0.2 (and also test $b_{\rm rot}=$0.4).

This model and the adopted parameters are motivated empirically by the dynamical Kennicutt-Schmidt relation \citep{Kennicutt:1998fv}, i.e., $\Sigma_{\rm SFR} \propto \Sigma_{\rm gas} \Omega$, where $\Omega$ is the orbital angular frequency.
This dynamical regulation of SF is expected in a variety of theoretical models, including: shear-driven cloud-cloud collisions \citep{Gammie:1991aa, Tan:2000lr,Tasker:2009dk, Li:2018aa}; spiral arm passage \citep{Wyse:1989aa}; and growth of large scale instabilities (e.g., \citep[][]{Elmegreen:1994aa, Wang:1994aa, Silk:1997uq} (see Section \ref{sec:dynamical} for further discussion).

\begin{figure}[t]
\epsscale{1.0}
\plotone{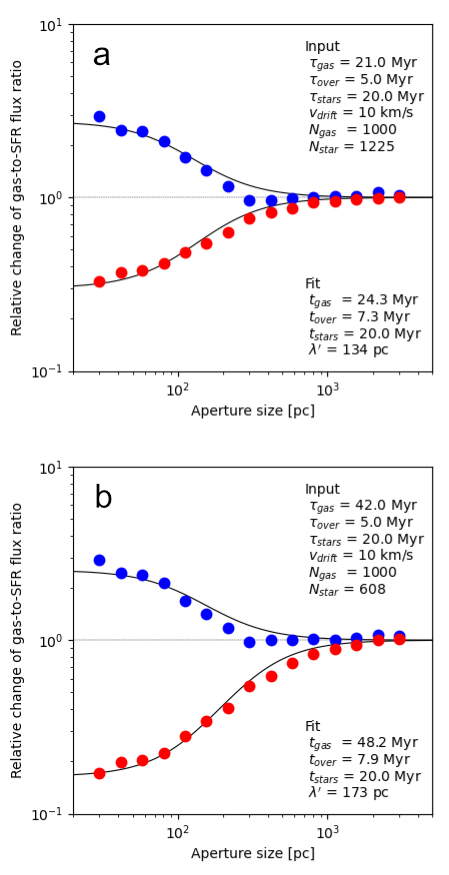}
\caption{
The same as Figure \ref{fig:uni} (uniform distribution), except for $p_{\rm SF}=\Delta t / (b_{\rm rot} t_{\rm rot})$,
i.e., $t_{\rm gas}=b_{\rm rot} t_{\rm rot}$ with the galactic rotation timescale $t_{\rm rot}$
and the efficiency of (a) $b_{\rm rot} = 0.2$ and (b) $b_{\rm rot} = 0.4$.}
\label{fig:collisions}
\end{figure}

Figure \ref{fig:collisions} shows the results of this model, which also produces TF diagrams. Since $t_{\rm gas}$ changes with galactic radius, we reference to an average $t_{\rm gas}$ over all clouds, which is shown in the top-right corners. We find that the TF fit provides an estimate for the average timescale of SF cycle in these immortal clouds.

\section{Discussion and Summary} \label{sec:discussion}

We have demonstrated that the TF analysis does not necessarily provide an estimate of cloud lifetimes.
Immortal clouds with occasional SF can produce a TF pattern if the stars drift away from their parental clouds.
The TF equations (eqs. \ref{eq:Bmodgas} and \ref{eq:Bmodstar}) can fit this TF pattern and return $t_{\rm gas}$ and $t_{\rm over}$.
This $t_{\rm gas}$ is \textit{not} the cloud lifetime, but a timescale of the SF cycle in the immortal clouds.
While the previous TF analyses were interpreted as implying short cloud lifetimes
\citep{Kruijssen:2019aa, Chevance:2020aa, Kim:2022aa}, we conclude that scenarios involving very long-lived clouds are also compatible.

As a result of the high molecular gas fraction and uneven distributions of clouds with and without SF over galactic disks (Section \ref{sec:intro}), we consider that scenarios for long-lived clouds are favored, especially in the samples used to define most of the dynamic range of the Kennicutt-Schmidt relation \citep{Kennicutt:1998fv}.

Our work has utilized a very simple toy model.
Some of the simplifications adopted here can easily be questioned.
However, the ``cloud lifetimes" derived with the TF analysis have often been granted without much considertaion on some other possibilities like the one presented in this work.
Therefore, the result that the TF analysis returns ``cloud lifetime" for immortal clouds is important for molecular cloud studies.

While the model in this work is intentionally kept simple, there are other potentially important physical processes in cloud evolution, e.g.:
various forms of stellar feedback; ram-pressure and tidal force of surrounding gas, including dense clouds; accretion of gas from the environment; and shearing effects in galactic disks, including spiral arms \citep[from numerous work in the literature:][]{Williams:1997aa, Dobbs:2013aa, Renaud:2013aa, Walch:2015aa, Kim:2016aa, Baba:2017um, Vazquez-Semadeni:2018aa, Kim:2018ab, Fujimoto:2019aa, Smith:2020aa, Kim:2021aa, Skarbinski:2023aa, Jeffreson:2023aa}.
Clouds can also fragment into smaller clouds or coagulate to larger clouds \citep[e.g., ][]{Scoville:1979lg, Tan:2000lr, Koda:2009wd, Tasker:2009dk, Dobbs:2013aa, Tan:2013aa, Inutsuka:2015aa, Kobayashi:2017aa, Kobayashi:2018aa}.
The validity of the TF analysis on clouds in such various evolutionary paths needs to be tested in the future.

If cloud evolution involves such loss or gain of mass, the definition of ``cloud lifetime" would need a clarification.
As a definition, we favor the full duration that the gas is molecular in molecular clouds, and hence is in a prerequisite condition for potential SF.
Most cloud studies stem from interests in SF, and it is more relevant to adopt the full duration that the gas can potentially form stars as cloud lifetimes \citep{Scoville:1979lg, Koda:2023aa}.

\subsection{The Stellar Drift Velocity -- $v_{\rm drift}$}

The stellar drift is a key assumption in this work (Section \ref{sec:drift}).
Unless stars move away from their parental clouds at some point in their evolution, the TF pattern is not produced.
Observationally, there are young stars, star clusters, and HII regions that are not closely associated with their parental clouds.

The stellar drift naturally arises in scenarios where SF is triggered by cloud-cloud collisions \citep{Scoville:1986aa, Tan:2000lr, Tasker:2009dk, Fukui:2014wd, Fukui:2021aa}, since the compressed regions of the clouds have significantly different velocities, of order the initial collision speeds, from the parts of the clouds not involved in the collision. Typical collision speeds in a shearing disk are $\sim 10$-$20\kmps$ \citep{Li:2018aa}.
Relative motion can also be induced due to stellar feedback, gas accretion, and/or cloud fragmentation and coagulation,
although some of these mechanisms would also tend to erode the molecular clouds.

\subsection{The Dynamical SF Model -- $b_{\rm rot}$}\label{sec:dynamical}

The SF model regulated by galactic rotation reproduces the TF pattern (Section \ref{sec:sfprob}).
This setup is motivated in particular by the model that SF is triggered by shear-driven cloud-cloud collisions in a galactic disk with an approximately flat rotation curve \citep{Gammie:1991aa, Tan:2000lr,Tasker:2009dk, Li:2018aa}.

We note that cloud-cloud collisions have been proposed by many authors to be an important process for triggering star formation \citep[e.g.,][]{Scoville:1986aa,Fukui:2014wd, Fukui:2021aa, Scoville:2023aa}, but have been criticized \citep[e.g., ][]{McKee:2007fk} for the timescale required between collisions. However, the shear-driven cloud collision model involves an effectively 2D monolayer of clouds in a galactic disk and predicts that collisions times can be relatively short, i.e., $b_{\rm rot} \sim 0.2$ \citep[and perhaps as short as 0.1 for the more massive clouds; ][]{Li:2018aa}.

The uneven spatial distribution of clouds with and without SF suggests that the trigger of SF occurs preferentially in spiral arms.
Mechanisms, such as cloud collisions involving long-lived clouds, are a natural way to help explain the observed spatial distribution, since the collisions are expected to be more common in spiral arms and less frequent in the interarm regions.

\subsection{The Scale Parameters -- $\lambda$ and $\lambda^{\prime}$} \label{sec:lambda}

KL14 and K18 introduced the $\lambda$ parameter ($=\lambda^{\prime}$ in our notation) as a spatial scale of small areas beyond which the areas become independent of each other in terms of SF (they called it an ``uncertainty principle").
Hence, $\lambda$ should measure a size scale that characterizes the scale of local SF events.
$\lambda$, if it could be measured, would constrain the physical processes that operate in SF and their characteristic size scales.

In the toy model in this work, the maximum stellar drift distance is the only spatial scale that characterizes the SF activity. However, the measured $\lambda$ (again, $\lambda^{\prime}$ in our notation) clearly depends on the global distribution.
For example, it is 234~pc for the uniform distribution (Section \ref{sec:caseAuni}), but it is 183~pc for the disk + spiral distribution (Section \ref{sec:caseAexpspi}), even though the other parameters beside the global distribution are virtually the same.
Hence, the $\lambda$ of KL14 and K18 (and our $\lambda^{\prime}$)
does not only trace the characteristic size scale of SF.
At the same time, we do find some sensitivity to local SF properties.
This can be see in Figure \ref{fig:comb_expspi} (left vs. right columns),
where the $\lambda$ (and our $\lambda^{\prime}$) increases with the maximum drift distance for larger $v_{\rm drift}$.

The key is the $\eta$ term (eqs. \ref{eq:Bmodgas} and \ref{eq:Bmodstar}) in $\lambda^{\prime}=\eta^{-1/2}\lambda$.
The $\lambda^{\prime}$ is what is measured in this work, as well as in \citet{Kruijssen:2019aa}, \citet{Chevance:2020aa}, and \citet{Kim:2022aa}.
This $\lambda^{\prime}$ parameter is not determined only by the local physical processes for SF (represented by $\lambda$), but depends on the global distribution of the SF activity (by $\eta$), such as the presence of spiral arms (see Sections \ref{sec:caseAuni} and \ref{sec:caseAexpspi}).
Therefore, the $\lambda^{\prime}$ can measure the size scale below which the global SF relations \citep{Kennicutt:2012aa} break down \citep{Schruba:2010fk, Onodera:2010fj}, but does not directly trace the local physical processes of SF \citep{Kruijssen:2019aa, Chevance:2020aa, Kim:2022aa}.

KL14 and K18 did not include this $\eta$ term in their TF equations (see eqs. 15, 16, C7, and C10 in K14 and eqs. 81 and 82 in K18).
In other words, they assumed that $\eta=1$, meaning that every small area of size $\lambda$ across the galactic disk and in an aperture either have a gas cloud or a star particle.
However, it is clear that observed galaxies are not entirely filled with molecular clouds nor SF regions (see Figure 1 of \citealt{Chevance:2020aa}).
In addition to the intrinsic lack of clouds and SF, $\eta$ could also be affected by the detection limits in observations.

\subsection{The Reference Timescale -- $t_{\rm star}$}

One of the difficulties of the short cloud lifetimes derived by previous TF studies \citep{Chevance:2020aa, Kim:2022aa} are the spatial distribution of molecular clouds with and without SF in spiral galaxies.
The clouds with SF are observed mostly along spiral arms, and those without SF are in the interarm regions.
The rotation timescale of galactic disks is on the order of 200~Myr, and hence, within the suggested short lifetimes of 5-20~Myr, the clouds cannot move from the interarm regions to spiral arms.
The immortal clouds encounter the same problem if the SF cycle timescale ($t_{\rm gas}$) is short.

We do not have a conclusive solution for this problem, but one possibility is an error in $t_{\rm star}$ (absolute timescale).
Recall the TF equations (\ref{eq:Bmodgas}) and (\ref{eq:Bmodstar}),
all the timescales appear as ratios (i.e., $t_{\rm gas}/t_{\rm tot}$, $t_{\rm over}/t_{\rm star}$, etc.).
Thus, the TF fit determines only those ratios.
Only when we know any of the timescales a priori, we can translate the ratios into absolute timescales.
In this work, we assumed that we \textit{know} $t_{\rm star}=20$~Myr (which is the input parameter).
However, when the TF analysis is applied to real observations,
we typically do not necessarily know the exact value of $t_{\rm star}$.

For example, in the case of the bottom-left panel of Figure \ref{fig:comb_expspi},
the TF analysis obtained the timescale ratios of $t_{\rm gas}/t_{\rm star}=59.0/20.0=2.95$ and $t_{\rm over}/t_{\rm star}=7.9/20.0=0.395$.
If we, by mistake, adopt $t_{\rm star}=5$~Myr as an absolute scale (instead of the input value of $t_{\rm star}=20$~Myr),
these ratios are translated to $t_{\rm gas}=$14.8~Myr and $t_{\rm over}=$2.0~Myr.
These are much shorter than the input values of $t_{\rm gas}=$50~Myr and $t_{\rm over}=$5.0~Myr,
but are similar to the values derived in many nearby galaxies
by \citet{Chevance:2020aa} and \citet{Kim:2022aa}
using the $t_{\rm star}=4.2$~Myr \citep{Haydon:2020ab}.
Their short $t_{\rm star}$ is estimated under an assumption that all stars in a star particle form simultaneously \citep{Haydon:2020ab}.
However, the observed age spreads in OB associations are known to be longer \citep[a few to a few tens of Myr; ][]{Wright:2023aa}.
The error in the adopted $t_{\rm star}$ may, entirely or partially, explain the apparently short $t_{\rm gas}$ derived in the previous studies.

There may be other potential sources of error in the TF analysis.
For example,  in observations, massive/luminous clouds exist preferentially in spiral arms \citep[e.g., ][]{Koda:2009wd, Colombo:2014aa, Rosolowsky:2021aa}.
The TF analysis uses the flux ratio (eq. \ref{eq:bobs}), and thus,
is biased toward the population of the brigter clouds in spiral arms.
The same bias can be introduced when the observations have a limited sensitivity and can detect clouds in  spiral arms, but not in interarm regions.
If the SF cycle is accelerated in spiral arms (as predicted, e.g,, by cloud collision models),
the TF analysis may return the short SF cycle timescale as $t_{\rm gas}$, while the interarm clouds can live long and remain dormant, passing from the interarm regions into spiral arms.
Such a possibility should be investigated in a future study.

\vspace{5mm}

\begin{acknowledgments}
JK thanks Fran\c{c}oise Combes for discussions and colleagues in Paris Observatory for hospitality during an extended stay on sabbatical.
We also thank the anonnymous referee, M\'elanie Chevance, Sarah Jeffreson, Masato Kobayashi, Diederik Kruijssen, Steve Longmore for valuable comments on the manuscript.
JK acknowledges support from NSF through grants AST-1812847 and AST-2006600. JCT acknowledges support from NSF through grants AST-2009674 and AST-2206450 and ERC through Advanced Grant 788829 (MSTAR).

\end{acknowledgments}

\bibliography{ms2.bbl}{}
\bibliographystyle{aasjournal}

\end{document}